\documentclass[prl,twocolumn,amsmath,amssymb,10pt,aps,superscriptaddress]{revtex4-1} 

\usepackage{feynmp}
\usepackage{amsmath}
\usepackage{graphicx}
\usepackage{rotating}
\usepackage{multirow}
\usepackage{latexsym}
\usepackage{textcomp}
\usepackage{verbatim}
\usepackage{color}
\usepackage{pict2e}
\usepackage{bm}
\usepackage{subfigure}
\usepackage{everysel}
\usepackage{keyval}
\usepackage{ragged2e}
\usepackage{dsfont}
\usepackage{amssymb}
\usepackage{enumitem}
\usepackage{pstricks}
\usepackage{mathtools}
\usepackage{hyperref}

\newcommand{\osum}{{%
    \setbox0\hbox{\circ}%
    \rlap{\hbox to \wd0{\hss\sum\hss}}\box0
}}

\newcommand{\beq}{\begin{equation}} 

\newcommand{\eeq}[1]{\label{#1} \end{equation}}

\newcommand{\cH}{\mathcal H}
\newcommand{\pd}{\partial}

\def\LBCO{La$_{2-x}$Ba$_x$CuO$_4$}

\def\C60{A$_x$C$_{60}$}

\def\HgCu3{HgCa$_2$Cu$_3$O$_{8+y}$}
\def\HgCu4{HgBa$_2$Ca$_3$Cu$_4$O$_{10+y}$}
\def\TlCu{Tl$_2$Ba$_2$CuO$_{6+\delta}$}
\def\TlCu3{Tl$_2$Ba$_2$Ca$_2$Cu$_3$O$_{10+y}$}
\def\TlCu4{Tl$_2$Ba$_2$Ca$_3$Cu$_4$O$_{12+y}$}

\def\BiCu3{Bi$_2$Sr$_2$Ca$_{2}$Cu$_3$O$_y$}

\def\8LSCO{La$_{1.88}$Sr$_{.12}$CuO$_4$}
\def\110LNSCO{La$_{1.5}$Nd$_{0.4}$Sr$_{0.1}$CuO$_{4}$}
\def\stage4LCO{La$_{2}$CuO$_{4+\delta}$}
\def\Y248{YBa$_2$Cu$_4$O$_8$}

\def\NbSe2{NbSe$_2$}
\def\TaSe2{TaSe$_2$}
\def\TiSe2{TiSe$_2$}
\def\NaCoOH2O{Na$_{0.3}$CoO$_{2y}$H$_2$O}
\def\MgB2{MgB${}_2$}

\def\URu2Si2{URu$_2$Si$_2$}

\def\Ba122{Ba(Fe$_{1-x}$Co$_x$)$_2$As$_2$}

\begin{document}

\title{Topological Pair-Density-Wave Superconducting States}   
\author{Gil Young Cho}
\affiliation{Department of Physics and Institute for Condensed Matter Theory, University of Illinois at Urbana-Champaign, 1110 
West Green Street, Urbana, IL  61801-3080, USA}
\author{ Rodrigo Soto-Garrido }
\affiliation{Department of Physics and Institute for Condensed Matter Theory, University of Illinois at Urbana-Champaign, 1110 
West Green Street, Urbana, IL  61801-3080, USA} 
\author{Eduardo Fradkin}         
\affiliation{Department of Physics and Institute for Condensed Matter Theory, University of Illinois at Urbana-Champaign, 1110 
West Green Street, Urbana, IL  61801-3080, USA} 
\affiliation{Kavli Institute for Theoretical Physics, University of California Santa Barbara, CA 93106-4030, USA}
\date{\today}    

\begin{abstract}
We show that the pair-density-wave (PDW) superconducting state emergent in  extended Heisenberg-Hubbard models in two-leg ladders is topological in the presence of an Ising spin symmetry and supports a  Majorana zero mode (MZM)   at an open boundary and at a junction with a uniform $d$-wave one-dimensional superconductor. Similarly to a conventional finite-momentum paired state,  the order parameter of the PDW state is a charge-$2e$ field with finite momentum. However, the order parameter here is a {\it quartic}  electron operator and conventional mean-field theory cannot be applied to study this state. We use bosonization to show that the 1D PDW state has a MZM at a boundary. This  superconducting state is an exotic topological phase supporting Majorana fermions with finite-momentum pairing fields and charge-$4e$ superconductivity. 
\end{abstract}

\maketitle

In the conventional theory of superconductivity, the Cooper pairs have zero center-of-mass momentum \cite{Bardeen-1957,schrieffer}. Fulde and Ferrell and independently Larkin and Ovchinikov showed that it is possible to have a superconducting (SC) state where the Cooper pairs have nonzero center-of-mass momentum in the presence of a uniform (Zeeman) magnetic field \cite{Fulde-1964, Larkin-1964}.
Evidence for  a nonuniform SC state has been 
found in 
some high-temperature superconductors. In these systems a nonuniform SC state appears with intertwined orders breaking translational symmetry, including spin-density wave (SDW) and charge-density wave (CDW) orders.\cite{kivelson-2003,vojta-2009,Fradkin-2012b,Fradkin-2010,fradkin-2014} An interesting example of the phase appears in the cuprate {\LBCO} (LBCO) \cite{Li-2007,tranquada-2008}. At $x=\tfrac{1}{8}$, the critical temperature $T_c$ of the uniform $d$-wave superconductivity is suppressed to near 4K. However, between 4K and 16K, where CDW and SDW orders are present, there is a quasi-two-dimensional SC phase, where CuO planes are superconducting but  the  material remains  insulating along the c-axis. This dynamical layer decoupling seen in LBCO (and  in LSCO in magnetic fields), i.e. an effective vanishing of the inter-layer Josephson coupling, can be explained if the CuO planes have pair-density-wave (PDW) SC order \cite{Berg-2007,Berg-2009}. This PDW SC state has been proposed as a natural competing state of the uniform $d$-wave SC state in the pseudogap regime
 \cite{berg-2008a,Berg-2009,fradkin-2014,lee-2014}.

In addition to the empirical evidence in cuprates \cite{fradkin-2014} there is
considerable 
evidence for the PDW state in microscopic models. Corboz {\it et.al.}\cite{corboz-2014} used iPEPS (infinite projected entangled pair-states) simulations and found strong evidence in the $t-J$ model (over a significant range of parameters) that the
PDW state, the uniform $d$-wave SC and and a coexistence SC state appear to have essentially degenerate in energy. Recently, two of us showed that in the weak coupling limit a PDW state is the ground state of a system in an electronic spin-triplet nematic phase \cite{Soto-Garrido-2014}. In 1D systems, a PDW state has been found in the spin-gap state of the Kondo-Heisenberg chain \cite{Berg-2010} and in an extended Hubbard-Heisenberg model on a two-leg ladder \cite{jaefari-2012}.

There is great interest in searching for  Majorana zero modes (MZM) in defects of topological SC states (vortices, junctions  and boundaries), ranging from one-dimensional wires with a proximity-induced superconductivity \cite{fidkowski-2011}, and chiral $p_x+ip_y$ SC states \cite{ivanov-2001}, to vortices in the SC surface of topological insulators \cite{alicea-2011, cheng-2011, fu-2008, fu-2010, cho-2012, alicea-2012}. Defects harboring MZM obey non-abelian statistics and are potential platforms for topological quantum computation, since the information is encoded non-locally and are immune to decoherence \cite{nayak-2008, alicea-2011}. In these cases, the SC states are uniform, and  the center-of-mass momentum of the Cooper pair is zero. Furthermore, the topological nature of the SC states can be understood from a weak coupling description of the states, {\it e.g.,} a mean-field theory, in which the superconductivity is encoded into the theory in terms of fermion bilinears.

In this work we show that MZM also appear  on one-dimensional (1D) systems in which the PDW SC state has been shown to be the ground state. As we will see below, the PDW SC states have {\em composite order parameters} which are quartic in the microscopic electronic degrees of freedom. Contrary to the conventional topological 1D SC states, these  1D PDW states cannot be described by the conventional Bogoliubov-de Gennes (BdG) mean field picture of superconductivity. For this reason it is not apparent how do these PDW states fit in the current classifications of 1D fermionic systems \cite{Kitaev-2009,ryu-2010,Fidkowski-2010}. The study of strongly correlated systems 
require the use of non-perturbative tools such as bosonization. Using  bosonization \cite{FradkinFieldTheory,Gogolin-book,Giamarchi-book} we show that the PDW SC state found in the two-leg ladder model and in the Kondo-Heisenberg model is  topological and supports a MZM at the end of the ladder. In this case, the MZM are associated with {\em solitons} of the spin sector of the PDW ground state which   are  a manifestation of the spin-charge separation of strongly correlated 1D fermionic systems.


\textbf{PDW States in Two-leg Ladder:} We start with the extended Hubbard-Heisenberg two-leg ladder model,  a physically relevant model for the study of cuprate superconductors, and demonstrate that the PDW SC state emergent from the model \cite{jaefari-2012} is topological in that it supports a MZM at the open boundary. With minor changes the same considerations apply to the spin gapped phase of (closely related) Kondo-Heisenberg chain \cite{Berg-2010}. In both systems, the PDW state has a spin gap and exhibits quasi-long range order only for order parameters which are quartic in electron fields (including an uniform charge $4e$ SC order parameter). In this highly non-mean-field SC state, all bilinears operators of the microscopic electrons have exponentially decaying correlations.

In a two-leg ladder, the local electron field $c_{a, \sigma, j}$ has the leg index $a \in \{1, 2 \}$, the site index $j \in {\mathbb Z}$, and the spin index $\sigma \in \{\uparrow, \downarrow \}$. In the presence of the inter-leg hopping, we first diagonalize the kinetic (hopping) term $H_{0}$ of the full two-leg ladder Hamiltonian $H=H_0+H_{\text{int}}$ using the bonding ($\eta = b$) and anti-bonding ($\eta = a$) basis states instead of the wire index   
\begin{align}
H_0 = \sum_{\eta=a,b} \sum_{j,\sigma}t_\eta \left( c^\dagger_{\eta,j,\sigma} c^{}_{\eta,j+1,\sigma} + \text{h.c.}\right),
\label{kinetic1}
\end{align}
where $t_\eta$ is the hopping parameter for the $\eta$-electron. In the low-energy limit, the kinetic term is  
\begin{equation}
	H_0 = \sum_{\eta,\sigma} \int dx (-i v_\eta)\left( R^\dagger_{\eta,\sigma}\partial_x R_{\eta,\sigma}
	-L^{\dagger}_{\eta,\sigma} \partial_x L^{}_{\eta,\sigma}\right), 
\label{kinetic2}
\end{equation}
where $v_\eta$ are the Fermi velocities for the two bands. The interaction terms can be rewritten in terms of charge and spin currents for the bonding and antibonding bands.\cite{jaefari-2012} We are interested in the case when the bonding band is at a rational filling and due to an umklapp operator  has a charge gap $\Delta_c>0$ (which for general filling requires a large enough nearest neighbor Coulomb interaction $V$). At low energy (compared to $\Delta_c$) the only charge degree of freedom is thus solely from the anti-bonding band. It is decoupled from the rest of the dynamics and its effective Hamiltonian is 
\begin{equation}
	{\cal H}_c = \frac{v_c}{2} \left( K_c(\partial_x\theta_c)^2+\frac{1}{K_c}(\partial_x\phi_c)^2 \right). 
\end{equation}
With the charge gap in the bonding band, the remaining interactions between the bonding band and the anti-bonding band  only involve their spin sectors. Thus the bonding band acts as the Heisenberg chain and couples to the anti-bonding electron through the Kondo coupling, and thus the model becomes identical to that of the Kondo-Heisenberg model.\cite{jaefari-2012, Berg-2010} The typical form of the interaction is $\sim J\bm{S}_a \cdot \bm{S}_b$, and the bosonized form of the Hamiltonian for the spin sector is \cite{wu-2003,jaefari-2012}
\begin{align}
 &{\cal H}_{s} = \frac{v_{s\pm}}{2} \left[ K_{s\pm}(\partial_x \theta_{s\pm})^2 + K^{-1}_{s\pm}(\partial_x \phi_{s\pm})^2 \right] \nonumber\\
  &+\frac{\cos(\sqrt{4\pi}\phi_{s+})}{2(\pi a)^2} \left[ g_{s1} \cos(\sqrt{4\pi}\phi_{s-}) + g_{s2} \cos(\sqrt{4\pi}\theta_{s-})\right],  
\label{spinsector}
\end{align}
where $\phi_{s\pm} = \frac{1}{\sqrt{2}}(\phi_{s,b} \pm \phi_{s,a})$ and similarly for $\theta_{s, \pm}$. (See the supplementary material A for a review of the two-leg ladder model of Ref.\cite{jaefari-2012} and bosonization details.) 

What is important here is that marginally relevant interaction term of Eq. \eqref{spinsector} drives the system into a regime in which the spin sector generally has a finite spin gap. In the spin gap phases (PDW and uniform SC), the operator  ``$\cos(\sqrt{4\pi} \phi_{s,+})$" in Eq. \eqref{spinsector} can be replaced by its expectation value $\mu_{\phi, s, +}$. With this approximation, valid deep inside the gapped phases, only the $(s,-)$ sector remains at  low energies and is subject to the potentials resulting from the second line of Eq. \eqref{spinsector}
\begin{align}
 &{\cal V}_{s} = \mu_{\phi, s, +}\left[ g_{s1} \cos(\sqrt{4\pi}\phi_{s-}) + g_{s2} \cos(\sqrt{4\pi}\theta_{s-})\right],  
\label{spinsector:potential}
\end{align}
In this regime, the resulting model has  two gapped phases: a commensurate PDW state with wave vector $Q=\pi$ with a stable fixed point at $(g_{s1}, g_{s2}) \to (-\infty, 0)$, and a uniform SC state $(g_{s1}, g_{s2}) \to (0, -\infty)$, with $K_{s,-} \to 1$. 
We are interested in the PDW phase described by the fixed point $(-\infty, 0)$ which has a two-fold degenerate ground state labelled by $\phi_{s,-}=0,\sqrt{\pi}$ (with the dual  field $\theta_{s,-}$  undefined). In this phase, the conventional SC and CDW order parameters have exponentially decaying correlations, but the PDW order parameter, represented by the composite operator (quartic in electron fields)
\begin{align}
O_\text{PDW} \sim \bigg(R^{\dagger}_{a}[i\sigma^{y}{\bm \sigma}] L^{*}_{a}\bigg)\cdot{\bm S}_b \sim (-1)^j\exp(i\sqrt{2\pi} \theta_{c}), 
\label{PDWOp}
\end{align}
  has power-law correlations due to the fluctuations of the surviving gapless charge mode $\theta_{c}$. The oscillatory prefactor reflects the short range commensurate order of the spin sector. We will show that this PDW phase is {\it topological} in that it supports a MZM at a junction with the uniform SC phase and  at an open boundary. 

The effective field theory of the spin sector $(s,-)$  at $K_{s,-} = 1$ that we  presented is  solved exactly in terms of a set of  new fermionic fields  \cite{FradkinFieldTheory,Gogolin-book,Giamarchi-book}  
\begin{align}
\mathcal{R} \sim  e^{-i\sqrt{\pi}(\phi_{s,-}-\theta_{s,-})}, \quad \mathcal{L} \sim   e^{i\sqrt{\pi}(\phi_{s,-}+\theta_{s,-})}, 
\label{spin:refermionization}
\end{align}
which are the {\it fermionic} excitations emergent at the low energies of the strongly coupled bosons described by Eq.\eqref{spinsector:potential}.  These (spinless) fermions  are unrelated to the microscopic electron appearing in Eq.\eqref{kinetic1} and Eq. \eqref{kinetic2}, and should be regarded as soliton states (or domain walls) 
that  interpolate between the two inequivalent ground states of the $\phi_{s,-}$ field! \cite{FradkinFieldTheory,Gogolin-book,Giamarchi-book}. In terms of the fermionic solitons, the potential of Eq. \eqref{spinsector:potential} becomes
\begin{align}
 &{\cal V}_{s} = M_{\text{uSC}} \mathcal{R}^{\dagger}\mathcal{L} + \Delta_{\text{PDW}} \mathcal{R}^{\dagger}\mathcal{L}^{\dagger} + h.c.,  
\label{spinsector:potential2}
\end{align}
with $M_{\text{uSC}} \sim \mu_{\phi,s,+}g_{s1}$ and $\Delta_{\text{PDW}} \sim \mu_{\phi,s,+} g_{s2}$. Hence, we  mapped the problem of the interacting $(s,-)$ spin sector into a problem of spinless fermions (solitons) with   masses $M_{\text{uSC}}$ and  $\Delta_{\text{PDW}}$. In Eq.\eqref{spinsector:potential2}, fermion number is not conserved but fermion parity, defined by
\begin{align}
(-1)^{\mathcal{N}_F} = (-1)^{\int dx~\left(\mathcal{R}^{\dagger}\mathcal{R} + \mathcal{L}^{\dagger}\mathcal{L}\right)} = e^{i \sqrt{\pi}\int dx ~\pd_{x}\phi_{s,-}}
\label{eq:fermion-parity}
\end{align} 
is conserved. The physical meaning of the fermion parity is the ${\mathbb Z}_{2}$ {\it spin parity}  which measures the parity of the relative change in the spin $S_{z}$ between the bonding and anti-bonding bands (see the supplementary material B). 

The potential of Eq.\eqref{spinsector:potential2} superficially resembles the pairing and CDW terms of a 1D spinless wire treated in the BdG mean field theory. However in the present case no mean field approximation was made (which strictly speaking does not hold in a 1D system). Instead, as we noted above, these spinless fermions are unrelated to the microscopic fermions of the ladder but are instead soliton excitations of this spin gap state.
Nevertheless, at this level, it is straightforward to identify the low-energy theory of Eq.\eqref{spinsector:potential2} with  the topological SC of class \textbf{D}, in which $M_{\text{uSC}}$ and $\Delta_{\text{PDW}}$ are interpreted as the conventional CDW and SC order parameters of spinless fermions (supplementary material A). Keeping this in mind, we now reveal the topological nature of the PDW state by showing that it has a MZM at a junction with the uniform SC state and at an open boundary. 

\textit{PDW-uSC junction}: We will now consider the case of a junction between a PDW state for $x>0$ and an  uniform ($d$-wave) SC for $x<0$. Roughly speaking, the junction between these two phases can be viewed as a ``phase transition  in real space'', instead of in parameter space. On the other hand, the quantum phase transition between the PDW and SC phases belongs to the Ising universality class.\cite{wu-2003,jaefari-2012} Across this phase transition the gap of a non-chiral Majorana fermion closes and opens up again. From this fact, we readily find that there should be a single Majorana fermion localized at the junction. 

To explicitly demonstrate this, we consider the junction configuration of $M_{\text{uSC}}(x)$ and $\Delta_{\text{PDW}}(x)$, i.e., they are the functions of the space $x$, such that $M_{\text{uSC}}(x)= 0$ for $x<0$ and non-zero for $x\geq 0$ and $\Delta_{\text{PDW}}$ is non-zero for $x\leq0$ but $0$ for $x>0$. We further rewrite the complex fermions $\mathcal{R}$ and $\mathcal{L}$ by the four Majorana fermions $\mathcal{R} = \eta_{R} + i \xi_R, \quad \mathcal{L} = \eta_{L} + i \xi_L$ whose Hamiltonian is 
\begin{align}
&{\cal H}_{s}  = -i v(\eta_R\pd_x\eta_R-\eta_L\pd_x\eta_L+\xi_R\pd_x\xi_R-\xi_L\pd_x\xi_L) \nonumber\\ 
&+ 2(M_{\text{uSC}}-\Delta_{\text{PDW}})i\eta_R\xi_L +2(M_{\text{uSC}}+\Delta_{\text{PDW}})i\eta_L\xi_R.   
\label{eq:majoranahamiltonian}
\end{align}
We find then that the Hamiltonian of Eq. \eqref{eq:majoranahamiltonian} is precisely the two copies of the Majorana fermions with the masses 
$(M_{\text{uSC}}-\Delta_{\text{PDW}})$ and $(M_{\text{uSC}}+ \Delta_{\text{PDW}})$. The fields $(\eta_L, \xi_R)$ will be always gapped with the size of the mass $|M_{\text{uSC}}+\Delta_{\text{PDW}}|>0$ near the junction at $x=0$. On the other hand, $(\eta_R, \xi_L)$ have the mass $M_{\text{uSC}}-\Delta_{\text{PDW}}$ which changes sign accross the junction and  vanishes  at $x \to 0$. We thus focus only on the fields $(\eta_R, \xi_L)$ for the low-energy physics of the junction
\begin{align}
&{\cal H}_{s} \approx -i v (\eta_R\pd_x\eta_R-\xi_L\pd_x\xi_L) + 2(M_{\text{uSC}}-\Delta_{\text{PDW}})i\eta_R\xi_L .   
\end{align}
This problem is equivalent to the Jackiw-Rebbi model \cite{jackiw-1976} and thus has a single MZM  exponentially localized at the junction.

\textit{Open Boundary}: We now show that the open boundary of the PDW state to the vacuum should also localize a single MZM. The PDW state is described by the potential of Eq.\eqref{spinsector:potential2} with $M_{\text{uSC}} =0 $ and non-zero $\Delta_{\text{PDW}}$. Then the low-energy Hamiltonian describing the $(s,-)$ sector is 
 \begin{align}
 &{\cal H}_{s} = (-i v)\left( \mathcal{R}^\dagger\partial_x \mathcal{R}
	-\mathcal{L}^{\dagger} \partial_x \mathcal{L}\right) + \Delta_{\text{PDW}} (\mathcal{R}^{\dagger}\mathcal{L}^{\dagger} + h.c.),  
\label{spinsector:refermion}
\end{align}
which is the low-energy theory of the spinless fermion exposed to the pairing, i.e., a topological SC in class \textbf{D}. A remarkable feature of this ``superconducting" spinless fermion state is that it has the dangling MZM at the open boundary.\cite{kitaev-2001, fidkowski-2011} In the free fermion class \textbf{D}, the MZM is solely protected by the fermion parity of the underlying microscopic electron, which leads to the ${\mathbb Z}_2$ classification. In Eq.\eqref{spinsector:refermion}, the fermions $(\mathcal{R}, \mathcal{L})$ are the fermionic solitons of the $(s,-)$ spin sector. Thus, the topological classification of the {\em spin sector} of the PDW state is {\it formally} equivalent to the class \textbf{D} with the fermion parity of the fermionic solitons Eq.\eqref{eq:fermion-parity}. Hence, in the presence of the $\mathbb{Z}_2$ symmetry generated by the parity, the refermionization of the $(s,-)$ sector is a theory  formally identical to that of the topological SC in class \textbf{D}. Thus it has the ${\mathbb Z}_{2}$ classification and supports a MZM at the open boundary as the free-fermion SC in class \textbf{D}, even though the microscopic  degrees of freedom of the ladder are  far from being free. 

The above results are derived from the refermionization of the $(s,-)$ sector at $K_{s,-} = 1$. However, the MZM has a topological origin and  is stable so far as the bulk of the PDW state is gapped and the associated Ising symmetry is respected. Thus, as far as the $(s,-)$ sector in the bulk is gapped and the Ising symmetry is respected, the MZM should be localized at the open boundary even with $K_{s,-} \neq 1$.\cite{cheng-2011} Thus this results hold in the entire PDW SC state and not only asymptotically.

Finally, in the PDW state, the spin sectors are gapped but the charge sector $\theta_c$ is gapless and decoupled. When the spin and the charge are completely decoupled and strictly separated, then the MZM, originated from the spin sector, is obviously stable. On the other hand, there are always irrelevant operators that mix charge and spin sectors. However the MZM couples only through the spin sector and the spin sector is gapped. Thus any term involving the spin sector, including the terms mixing spin and charge sectors, has exponentially decaying correlation length and so the MZM is exponentially localized at the junction or boundaries. Thus the MZM is stable despite of the gapless charge sector.

\textbf{Two PDW Ladders:} Because of the ${\mathbb Z}_{2}$ nature of the MZM in the PDW state, one may naively expect that the system of the two coupled PDW ladders should be trivial. We will show that it is not the case because of the charge sector, and that there can be a MZM from the charge sector though the MZM from the spin sectors are actually gapped out.  

Indeed, the charge sector of each ladder remains gapless in the PDW phase of Eq.\eqref{spinsector}, and there are two {\it local} order parameters exhibiting power-law correlations \cite{jaefari-2012}. The first is the PDW order parameter $O_\text{PDW}$ of Eq.\eqref{PDWOp}, and the other is the CDW order parameter at momentum $2k_{F,a} + \pi$ (where $k_{F,a}$ is the Fermi wave vector of the antibonding band of the ladder)
\begin{align}
O_{\text{CDW}} (x) \sim \bigg(R^{\dagger}_{a} {\bm \sigma}L_{a}\bigg) \cdot {\bm S}_{b} \sim \exp(i\sqrt{2\pi} \phi_{c}).
\end{align}
Let us consider now a system of two coupled two-leg ladders.
Due to the spin gap in the PDW states, the single particle tunneling and any spin-spin coupling between the ladders is irrelevant. The only remaining local perturbations at the decoupled fixed point involve $O_{\text{PDW}}$ and $O_{\text{CDW}}$ (see Ref.\cite{jaefari-2012})
\begin{align}
\delta H &= - {\mathcal J} O^{\dagger}_{1, \text{PDW}}O_{2, \text{PDW}} - g O^{\dagger}_{1, \text{CDW}} O_{2, \text{CDW}} +h.c., \nonumber\\ 
&= -{\mathcal J} \cos(\sqrt{4\pi} \theta_{c, -}) - g \cos(\sqrt{4\pi} \phi_{c,-}), 
\label{twoladder}
\end{align}
where $1,2$ label the two ladders and $\phi_{c,-} = \tfrac{\phi_{c,1} - \phi_{c,2}}{\sqrt{2}}$ (similarly for $\theta_{c,-}$). Despite of the simple appearance of Eq.\eqref{twoladder}, these terms are actually octets in electron fields (!) and are usually ignored in  lattice model Hamiltonians. However, all the local quartic terms, e.g. $J \bm{S}_1 (x) \cdot \bm{S}_2 (x)$, in the Hamiltonian are irrelevant at the PDW phase, and thus the terms in Eq.\eqref{twoladder} are the most relevant perturbations at this fixed point. 

The Hamiltonian of Eq.\eqref{twoladder} can, again, be mapped to Eq.\eqref{spinsector:potential2} by refermionization of the $(c,-)$ charge sector. Thus, when the Josephson coupling ${\mathcal J}$  is relevant (and flows to infinity), there will be a  MZM from the $(c,-)$ sector. On the other hand, the Majorana fermions from the spin sector will be generically gapped out by the local spin-spin interactions between the ladders. From this example, we see that the naive expectation, that the coupling of the two topological SC wires should result a trivial state, may not be true and the coupling may result in a surprising topological state if each wire contains gapless modes (here the gapless sector is the charge sector). Based on the observation on the two coupled PDW ladders, we can treat quasi-one-dimensional systems in which many PDW wires are stacked and coupled each other as done in the supplementary material C. In the quasi-one-dimensional states,\cite{jaefari-2010, cho-2012b, ran-2010, teo-2013, kivelson-1998, emery-2000} there are various weak topological phases of the charge and spin sectors and MZM at lattice defects.


In spite being topologically trivial, the CDW phase $g \rightarrow \infty$ in Eq.\eqref{twoladder} is not a usual insulating state. The charge sector $(c,-)$ of the two-ladder system is in the CDW phase, which implies that $O_{a, \text{PDW}}, a= 1,2$ has an exponentially decaying correlations. However, in this phase a uniform $4e$ SC order parameter $\Delta_{4e} \sim O_{1,\text{PDW}}O_{2, \text{PDW}}$ has the power-law correlation since the $(c,+)$ sector is gapless. 


\textbf{Conventional PDW state:} In the above, we have considered a particular PDW state emergent in a strongly-coupled Heisenberg-Hubbard two-leg ladder, in which the PDW order parameter of Eq.\eqref{PDWOp} has the commensurate momentum. Here we consider a more conventional finite-momentum SC state emerging from an one-dimensional system with the spin-rotation, time-reversal, and translation symmetries. Thus we consider the model with the four fermi points at $k = \pm k_{F,1}$ and $k = \pm k_{F,2}$, and each fermi point is doubly degenerate due to the electronic spin $\sigma = \uparrow, \downarrow$, 
\begin{align}
\Psi_{a,\sigma}(x) \sim e^{i k_{F,a}x} R_{a, \sigma} (x) + e^{-ik_{F,a}x} L_{a, \sigma}(x), 
\end{align}
with $a \in \{1,2\}$. We consider a {\it phenomenological} effective local attractive interaction
\begin{align}
&\delta H = \frac{V_{0}}{4}[\Psi^{*}_{1, \alpha}(i\sigma^{y})^{\alpha\beta}\Psi^{*}_{2, \beta}]   [\Psi_{1, \lambda}(i\sigma^{y})^{\lambda\delta}\Psi_{2, \delta}], \nonumber\\ 
&= -\frac{V_{0}\cos(\sqrt{4\pi}\theta_{s,-}) }{(2\pi a)^2} \bigg(\cos(\sqrt{4\pi} \phi_{s,+}) + \cos(\sqrt{4\pi}\phi_{s,-})\bigg), 
\label{CPDW}
\end{align}
which is identical to Eq.\eqref{spinsector} except the irrelevant term $\sim \cos(\sqrt{4\pi}\theta_{s,-}) \cos(\sqrt{4\pi}\phi_{s,-})$.\cite{Berg-2010} The model \eqref{CPDW} is simply a model of the two {\it inequivalent}1D spin-1/2 wires coupled by an attractive interaction. Now when the pairing potential becomes relevant, i.e., deep in the SC phase, we can first ignore the irrelevant term $\sim \cos(\sqrt{4\pi}\theta_{s,-}) \cos(\sqrt{4\pi}\phi_{s,-})$ in \eqref{CPDW} and replace $\langle \cos(\sqrt{4\pi}\phi_{s,+}) \rangle$ by its expectation value $\mu_{\phi, s,+}$. Then following the discussion in the PDW state of the two-leg ladder, we refermionize the $(s,-)$ sector and find that there will be a MZM at the open boundary if the associated Ising symmetry is present as in the two-leg ladder PDW state case. Here the PDW order parameters $O_{\text{PDW}}(x) \sim [\Psi^{\dagger}_{a}i\sigma^{y}{\bm \sigma}\Psi^{*}_{a}] \cdot {\bm S}_{b}, a \neq b$ will develop a power-law correlations. From this example, we see that the commensurability of the PDW order parameter of Eq.\eqref{PDWOp} is not important, and that the emergence of the MZM may be more general than the particular PDW model discussed in Eq.\eqref{spinsector}.    

\begin{table}[tb!]
\begin{tabular} {c|c|c}
\hline
Model & Phase&  Sectors  \\ \hline
Two-leg Hubbard ladder & PDW  & (s, -)  \\ \hline
Two-coupled PDW Ladders & PDW & (c, -) \\ \hline
Two-coupled PDW Ladders & CDW &  none \\ \hline
Conventional PDW state & PDW  & (s, -)  \\ \hline
\end{tabular}
\caption{Topological Phases with MZM.}
\label{tab:result}
\end{table}

\textbf{Conclusions:} In this Letter, we  discussed the emergence of the MZM in the PDW state of two-leg ladders. In this state, the  PDW order parameter is quartic in  electron operators, and  the topological nature cannot be studied within  mean-field theory. Using  bosonization, we showed  that the state is topological, and supports a MZM at the open boundary. The main results are summarized in the table \ref{tab:result}. The MZM discussed in this Letter emerges from the fermionic solitons  of the spin or charge sectors, and are not simply related to the  microscopic electronic degrees of freedom. 
This MZM is a feature of the soliton spectrum of the spin sector of the two-leg ladder (and of the charge sector of two coupled two-leg ladders) which should dominate the low energy response in an (idealized) {\em electron} tunneling experiment.
The robust two-fold degeneracy coming from the Majorana fermions should appear in the entanglement spectrum as the two-fold degeneracy of the lowest eigenvalue of the entanglement Hamiltonian \cite{stoudenmire-2011,turner-2011}.

\begin{acknowledgments}
We thank J.C.Y. Teo, I. H. Kim, and M. Cheng for useful discussions. EF thanks the KITP (and the Simons Foundation) and its IRONIC14 program for support and hospitality. This work was supported in part by the NSF grants DMR-1064319 (GYC) and DMR 1408713 (EF) at the University of Illinois,  PHY11-25915 at KITP (EF),  DOE Award No. DE-FG02-07ER46453  (RSG) and  Program Becas Chile (CONICYT) (RSG).  
\end{acknowledgments}

\end{document}


\title{Supplemental Material for ``Topological Pair-Density-Wave Superconducting States"}
\author{Gil Young Cho, Rodrigo Soto-Garrido and Eduardo Fradkin}
\affiliation{Department of Physics and Institute for Condensed Matter Theory, University of Illinois, 1110 W. Green St., Urbana IL 61801-3080, U.S.A.}
\date{\today}
\maketitle
\appendix

\section{Class D topological superconductor Wire and Bosonization} 
In this supplementary appendix, we review and study the BdG theory of the topological superconductor in class \textbf{D} of the spinless fermion wire rather carefully in the bosonized langauge.\cite{fidkowski-2011} We carefully examine the number of the degenerate ground states in terms of the bosonic variables.\cite{fidkowski-2011} Furthermore, we will take this supplementary appendix as the chance to review the bosonization convention used throughly in this Letter.\cite{FradkinFieldTheory} 

The Hamiltonian for a system of spinless fermions exposed to a p-wave SC pairing is given by:
\begin{align}
 H=-t\sum_j\left(c^{\dagger}_{j}c^{}_{j+1}+\text{h.c.}\right)-\mu\sum_jc^{\dagger}_{j}c^{}_{j}-\sum_j\left(\Delta^*c^{}_{j}c^{}_{j+1}+\text{h.c.}\right)
 \label{kitaev}
\end{align}
In the continuum limit and at the low energy, we write:
\begin{equation}
 	\frac{1}{\sqrt{a}}c_{j} \rightarrow\psi(x)= R(x)e^{ ik_{F}x} + L(x)e^{-ik_{F}x},
\end{equation}
with the ultra-violet cut-off `$a$'. Using the stardard bosonization technique \cite{FradkinFieldTheory} we define the bosonic fields via its relation with the slow fermions:
\begin{align}
 R(x)=\frac{1}{\sqrt{2\pi a}}e^{-i\sqrt{\pi}(\phi-\theta)}, \quad L(x)=\frac{1}{\sqrt{2\pi a}}e^{i\sqrt{\pi}(\phi+\theta)},
\end{align}
where the fields $\phi$ and $\theta$ are dual to each other. It is apparent that the fermion fields $ R(x),L(x)$ are invariant under $\phi \rightarrow \phi + 2\sqrt{\pi}$, and this sets the compactness of the bosonic variable $\phi$. The same holds for $\theta$. Thus, we find that $(\phi, \theta)$ labels the same state as $(\phi +2\sqrt{\pi}, \theta)$, {\it i.e.,} we have 
\begin{align}
(\phi, \theta) \sim (\phi+2\sqrt{\pi}, \theta) \sim (\phi, \theta+2\sqrt{\pi}). 
\end{align}
Furthermore the bosonic variables $(\phi, \theta)$ satisfy the standard equal-time canonical commutation relation
\begin{equation}
[\phi(x),\partial_{x'} \theta(x')]=i\delta(x-x'),
\label{commutator}
\end{equation}
which identifies $\Pi=\partial_x \theta$ as the canonical momentum of $\phi$. The associated currents are given by:
\begin{equation}
 j_0=\frac{1}{\sqrt{\pi}}\pd_x\phi\qquad\text{and}\qquad  j_1=-\frac{1}{\sqrt{\pi}}\pd_x\theta
\end{equation}
In this simple case (spinless fermions), in addition to the usual Luttinger liquid terms we consider two extra terms. One of them is a CDW mass term (at the momentum $\pm 2k_F$) given by 
\begin{equation}
 R^{\dagger}L + h.c.= \frac{1}{\pi a}:\cos(\sqrt{4\pi}\phi) :, 
\end{equation}
and a SC term given by:
\begin{equation}
 R^{\dagger}L^{\dagger}+h.c.=\frac{1}{\pi a}:\cos(\sqrt{4\pi}\theta):,  
\end{equation}
with a properly defined normal ordering $:A: = A - \langle A \rangle$ for the vertex operators. In this Letter, all the cosine terms are normal ordered implicitly. So, in this mean-field level, the Hamiltonian in eq. \eqref{kitaev} becomes (supplemented by the CDW term):
\begin{align}
 \cH=\frac{v_F}{2} \left( K(\partial_x\theta)^2+\frac{1}{K}(\partial_x\phi)^2 \right) +M\cos(\sqrt{4\pi}\phi)+\Delta\cos(\sqrt{4\pi}\theta)
 \label{boskitaev}
\end{align}
Now for $\Delta<0$ and being relevant ({\it i.e.}, $\Delta \rightarrow - \infty$), we find that
\begin{equation}
 \ev{\cos(\sqrt{4\pi}\theta)}=1,
 \label{valtheta}
\end{equation}
minimizes the energy. Taking into account that the compactification radius of $\theta$, we have the two possible values for $\theta =0 , \sqrt{\pi}$. Let us see now how these two values of $\theta$ are related.
Defining the charge operator as:
\begin{equation}
 Q=\int dx'j_0(x')=\int dx'\frac{1}{\sqrt{\pi}}\pd_{x'}\phi(x')
\end{equation}
and using the commutator eq. \eqref{commutator} we have that:
\begin{equation}
 e^{i\pi Q}\theta e^{-i\pi Q}=\theta+\sqrt{\pi} 
\end{equation}
Thus we notice that the unitary operator $e^{i\pi Q}$ corresponds to the fermion parity operator $(-1)^{N_F}$ and that the two values of $\theta=\{0,\sqrt{\pi}\}$ are mapped to each other by the fermion parity.\cite{fidkowski-2011} Requiring the ground states to carry the definite fermion parity, we find that 
\begin{equation}
(-1)^{N_F} \frac{|0 \rangle \pm | \sqrt{\pi} \rangle}{\sqrt{2}} = \pm \frac{| 0 \rangle \pm |\sqrt{\pi} \rangle}{\sqrt{2}},  
\end{equation}
where we have defined a {\it kat} $|\theta_{0} \rangle $ as the {\it eigenstate} of the bosonic field $\theta$ 
\begin{equation}
\theta |\theta_{0} \rangle = \theta_{0} |\theta_{0} \rangle.
\end{equation}
Then we see that the two ground states $\frac{|0 \rangle \pm |\sqrt{\pi} \rangle}{\sqrt{2}}$ have different fermion parities.\cite{fidkowski-2011} As far as the fermion parity is conserved, there is no matrix element in the Hamiltonian connecting the two states. Hence, the two-fold degeneracy is exact in the thermodynamic limit in which the system size $L \rightarrow \infty$. This is not the case for the CDW state, where the two-fold degeneracy of the ground states of the term $\sim M \cos(\sqrt{4\pi} \phi), M \rightarrow -\infty$ in the Hamiltonian is not protected by the fermion parity. Thus the SC state is topological, but the CDW state is not.

\section{Two-leg Ladder model and ${\mathbb Z}_2$ fermion parity in PDW state} 
In this supplementary appendix, we review the Heisenberg-Hubbard two-leg ladder model (which supports the PDW state as the ground state) and provides the understanding of ${\mathbb Z}_2$ fermion parity of the fermionic soliton emergent from the refermionization of the $(s,-)$ sector in the PDW ladder. 

We start by reviewing some of the main results on the Hubbard-Heisenberg model in a two leg ladder. In the following we closely follows 
Ref. [\onlinecite{jaefari-2012}]. The Hamiltonian $H=H_0+H_{\text{int}}$ is given by:
\begin{align}
\label{kineticterm}
 H_0=-t\sum_{a,j,\sigma} \left( c^{\dagger}_{a,j,\sigma}c_{a,j+1,\sigma}+h.c. \right)-t_{\perp}\sum_{j,\sigma} \left( c^{\dagger}_{1,j,\sigma}c_{2,j+1,\sigma}+h.c. \right)
\end{align}
where $t$ and $t_{\perp}$ are the intraleg and interleg hopping amplitudes, $a = 1,2$ is the leg index, and $j$ is the lattice site index. 
The interaction part of the $H$ is given by:
\begin{align}
H_\text{int}= & U\sum_{a,j} n_{a,j,\up}n_{a,j,\dn} + V_\parallel \sum_{a,j} n_{a,j}n_{a,j+1}+ V_\perp  \sum_{j} n_{1,j}n_{2,j} +V_d  \sum_{j} \left(n_{1,j}n_{2,j+1}+n_{1,j+1}n_{2,j} \right)\nonumber\\
&+ J_{\parallel} \sum_{a,j} \vec S_{a,j} \cdot \vec S_{a,j+1} + J_\perp \sum_{j} \vec S_{1,j} \cdot \vec S_{2,j}+ J_d  \sum_{j} \left( \vec S_{1,j} \cdot \vec S_{2,j+1}+\vec S_{1,j+1} \vec S_{2,j} \right)
\label{lattice-model}
\end{align}
where $U$ is the on-site Hubbard repulsion, $V_\parallel$, $V_\perp$ and $V_d$ are the  nearest and next-nearest neighbor Coulomb repulsions, 
and $J_\parallel$, $J_\perp$ and $J_d$ are the nearest and next-nearest neighbor exchange interactions. The kinetic term eq. \eqref{kineticterm}
can be diagonalized. In this diagonal basis the kinetic term takes the simples form:
\begin{align}
	H_0 = \sum_{\eta=a,b} \sum_{j,\sigma}t_\eta \left( c^\dagger_{\eta,j,\sigma} c^{}_{\eta,j+1,\sigma} + \text{h.c.}\right)
\end{align}
where $a$ and $b$ label the antibonding and bonding bands and $t_\eta=t \pm t_\perp$ for $\eta=b,a$ respectively. In the continuum limit and at low
energies we can write:
\begin{equation}
 	\frac{1}{\sqrt{a}}c_{\eta,j,\sigma} \rightarrow R_{\eta,\sigma}(x)e^{ ik_{F\eta}x} + L_{\eta,\sigma}(x)e^{-ik_{F\eta}x},
\end{equation}
where $R$ and $L$ are right- and left-moving components of the electron field, $x=ja$ is the position, and $a$ is the lattice constant. 
In this limit, the kinetic term of the Hamiltonian takes the standard continuum form 
\begin{equation}
	H_0 = \sum_{\eta,\sigma} \int dx (-i v_\eta)\left( R^\dagger_{\eta,\sigma}\partial_x R_{\eta,\sigma}
	-L^{\dagger}_{\eta,\sigma} \partial_x L^{}_{\eta,\sigma}\right)
\end{equation}
where $v_\eta$ are the Fermi velocities for the two bands. This low-energy fermion theory can be bosonized through 
\begin{align}
 R_{\eta,\sigma}=\frac{F_{\eta,\sigma}}{\sqrt{2\pi a}}e^{-i\sqrt{\pi}(\phi_{\eta,\sigma}-\theta_{\eta,\sigma})}, \quad 
 L_{\eta,\sigma}=\frac{F_{\eta,\sigma}}{\sqrt{2\pi a}}e^{i\sqrt{\pi}(\phi_{\eta,\sigma}+\theta_{\eta,\sigma})}
\end{align}
where $\eta,\sigma$ labels the band and the spin polarization ($j=b,a$ and $\sigma=\up,\dn$). The Klein factors, $F_{\eta,\sigma}$, ensure that the
fermions with  different labels anti-commute $\{F_{\eta,\sigma},F_{\eta',\sigma'} \}=\delta_{\eta,\eta'}\delta_{\sigma,\sigma'}$. We now define the charge and spin bosonic fields as:
\begin{equation}
 \phi_{\eta,c/s}=\frac{\phi_{\eta,\up}\pm\phi_{\eta,\dn}}{\sqrt{2}}
\end{equation}
Or more explicitly:
\begin{align}
 \phi_{b,c}=\frac{1}{\sqrt{2}}\left(\phi_{b,\up}+\phi_{b,\dn} \right), ~
 \phi_{b,s}=\frac{1}{\sqrt{2}}\left(\phi_{b,\up}-\phi_{b,\dn} \right), \text{ and  }
 \phi_{a,c}=\frac{1}{\sqrt{2}}\left(\phi_{a,\up}+\phi_{a,\dn} \right), ~
 \phi_{a,s}=\frac{1}{\sqrt{2}}\left(\phi_{a,\up}-\phi_{a,\dn} \right).
 \label{eq:phis}
\end{align}
Then the interaction term can be rewritten fully in terms of charge and spin bosonic fields for the bonding and antibonding bands. The explicit expression can be found in Ref. [\onlinecite{jaefari-2012}]

From now on we will consider the case when the bonding band is half filled and its Fermi wave vector is $k_{Fb}=\pi/2$. 
In this case there is a charge gap $\Delta_c$ in the bonding band. 
This case is simpler, since at low energies (at least small compare to $\Delta_c$) we can assume that the charge degrees of freedom on 
bonding-band $b$ are effectively frozen and play no roll in the low energy limit of the remaining degrees of freedom. 
In addition, the SC terms between the bands are irrelevant (since those produce a net charge transfer between the bands). 
In this limit the only charge degrees of freedom are those of the anti-bonding band $a$, and are decoupled from the rest of the dynamics. 
In its bosonized form the effective Hamiltonian density for the charge sector involves the Bose field $\phi_c$ and its dual field $\theta_c$ 
for the anti-bonding band $a$ only, which is given by (the usual Luttinger liquid (LL) theory),
\begin{equation}
	{\cal H}_c = \frac{v_c}{2} \left( K_c(\partial_x\theta_c)^2+\frac{1}{K_c}(\partial_x\phi_c)^2 \right)
	\label{eq:Hc} 
\end{equation}
On the other hand, the Hamiltonian for the spin sector is given by (again, for details see Ref. [\onlinecite{jaefari-2012}]):
\begin{align}
 &{\cal H}_{s} = \frac{v_{s\pm}}{2} \left[ K_{s\pm}(\partial_x \theta_{s\pm})^2 + K^{-1}_{s\pm}(\partial_x \phi_{s\pm})^2 \right] +\frac{\cos(\sqrt{4\pi}\phi_{s+})}{2(\pi a)^2} \left[ g_{s1} \cos(\sqrt{4\pi}\phi_{s-}) + g_{s2} \cos(\sqrt{4\pi}\theta_{s-})\right]
 \label{eq:Hs}
\end{align}
where $\phi_{s\pm} = \frac{1}{\sqrt{2}}(\phi_{s,b} \pm \phi_{s,a})$ (and the same for $\theta_{s\pm}$). Using this Hamiltonian Jaefari and Fradkin 
\cite{jaefari-2012} showed that there are three different phases. One of the them is a C1S2 Luttinger state (one charge and two spin gapless modes). The second phase corresponds to a uniform superconducting state (coexisting with a charge density wave (CDW) state). Finally, and more interesting, there is PDW SC phase. An interesting feature of this phase is that eventhough the SC order parameters and the CDW order parameter are short-range, the PDW OP defined by:
\begin{align}
	&O_\text{PDW} = \vec\Delta_a\cdot\vec N_b=\frac{1}{2(\pi a)^2} \cos(\sqrt{2\pi}\phi_{c,b})e^{-i\sqrt{2\pi}\theta_{c,a}} \times \left[2 \cos(\sqrt{4\pi}\theta_{s-}) +\cos(\sqrt{4\pi}\phi_{s-}) - \cos(\sqrt{4\pi}\phi_{s+})\right].
					\label{O-PDW}
\end{align}
presents power law correlations:
\begin{equation}
  	\ev{O^{}_\text{PDW}(x)O^{\dagger}_\text{PDW}(0)} \sim {\cal C}^2_c {\cal C}^2_s |x|^{-2/K_{c,a}}
\end{equation}
This operator, being quartic in fermionic operator, differs from the usual treatment at the mean field (MF) level. 
In that case, the $O^{}_\text{PDW}$ is a bilinear and the system can usually study at the MF level. 

Now we assume that we are deep in the PDW state, i.e., $g_{s2} \to -\infty$ and $g_{s1} \to 0$ in Eq.\eqref{eq:Hs}. Deep inside the PDW state, the ground state satisfies $\langle \cos(\sqrt{4\pi} \phi_{s,+}) \cos(\sqrt{4\pi} \theta_{s,-}) \rangle = 1$, and thus $\langle \cos(\sqrt{4\pi} \phi_{s,+}) \rangle  = \mu_{\phi, s,+} = \pm1$. Hence, deep inside the PDW phase, we find that the Hamiltonian Eq.\eqref{eq:Hs} for the spin sector becomes 
\begin{align}
 &{\cal H}_{s} \approx \frac{v_{s-}}{2} \left[ K_{s-}(\partial_x \theta_{s-})^2 + K^{-1}_{s-}(\partial_x \phi_{s\pm})^2 \right] + \mu_{\phi, s,+} \cos(\sqrt{4\pi}\theta_{s-}) 
 \label{eq:Hs2}
\end{align}
On the other hand, the Luttinger parameter $K_{s-}$ for the $(s,-)$ spin sector flows to $1$ asymptotically. Thus the low-energy Hamiltonian Eq.\eqref{eq:Hs2} can be refermionized and solved exactly by the refermionization as done in the main text. So we introduce the spinless fermion fields \cite{FradkinFieldTheory,Gogolin-book,Giamarchi-book}  
\begin{align}
\mathcal{R} \sim  e^{-i\sqrt{\pi}(\phi_{s,-}-\theta_{s,-})}, \quad \mathcal{L} \sim   e^{i\sqrt{\pi}(\phi_{s,-}+\theta_{s,-})}, 
\label{spin:refermionization}
\end{align}  
and we rewrite the Hamiltonian Eq.\eqref{eq:Hs2} in terms of this fermionic solitons as done in the main text.  
\begin{align}
{\cal H}_{s} = (-i v)\left( \mathcal{R}^\dagger\partial_x \mathcal{R} -\mathcal{L}^{\dagger} \partial_x \mathcal{L}\right) + \Delta_{\text{PDW}} (\mathcal{R}^{\dagger}\mathcal{L}^{\dagger} + h.c.),  
\label{spinsector:refermion}
\end{align}
with which we have identified $\Delta_{\text{PDW}} \sim \mu_{\phi,s,+} g_{s2}$. Then it is precisely the same as the BdG Hamiltonian of the class \textbf{D} topological SC Eq.\eqref{boskitaev}. The two-fold degenerate ground states in the PDW state is identified by the classical value of $\theta_{s,-} \in \{0, \sqrt{\pi} \}$ (for $\Delta_{\text{PDW}}<0$) and the two ground states are mapped each other by the fermion parity operator of the fermionic solitons 
\begin{align}
(-1)^{\mathcal{N}_F} = (-1)^{\int dx~\left(\mathcal{R}^{\dagger}\mathcal{R} + \mathcal{L}^{\dagger}\mathcal{L}\right)} = e^{i \sqrt{\pi}\int dx ~\pd_{x}\phi_{s,-}}
\end{align}
To understand the physical meaning of this parity operator, we rewrite it as 
\begin{align}
 (-1)^{N_F} = (-1)^{Q_{s,-}}, ~Q_{s,-}= \frac{1}{\sqrt{\pi}}\int dx \pd_x\phi_{s-}.
\end{align} 
The parity $ (-1)^{Q_{s,-}}$ measure the parity of the {\it relative change} in the spin $S_{z}$ between the bonding and anti-bonding bands. More properly, $Q_{s,-} = S_{z, a} - S_{z,b}$ is the generator of the spin rotational symmetry around the $z$-axis, {\it i.e.,} $U_{\delta \phi} = \exp(i \delta \phi Q_{s,-}), \delta \phi \in [0, 2\pi]$ rotates the spin around $z$-axis by $+ \delta \phi$ for the bonding electron and $-\delta \phi$ for the anti-bonding electron. Thus the parity $(-1)^{Q_{s,-}}$ is in fact an Ising-symmetry operation. With this understanding in hand, we imagine that we start with the ground state satisfying $N_{\uparrow, b} = N_{\downarrow, b} \in {\mathbb Z}$ and $N_{\uparrow,a} = N_{\downarrow,a} \in {\mathbb Z}$. Then we find that 
\begin{equation}
Q_{s,-} = \frac{1}{2} \bigg( (N_{\uparrow,b} - N_{\downarrow,b}) - ( N_{\uparrow,a} - N_{\downarrow,a}) \bigg) \in {\mathbb Z},
\end{equation} 
in the presence of the Ising symmetry, and the parity $ (-1)^{Q_{s,-}}$ maps a value of $\theta_{s-}=\{0,\sqrt{\pi}\}$ to the other.

\section{Quasi-one-dimensional PDW states} 
In this supplementary material, we consider a quasi-one-dimensional system in which we stack the two-leg ladders in a PDW state. 
Because the charge sector is completely decoupled from the spin sector in the PDW state, we first discuss the phases of the charge sectors. The leading allowed  local perturbations at the decoupled fixed point are 
\begin{align}
\delta H =  -{\mathcal J} \sum_{\langle i,j\rangle} & \cos \bigg(\sqrt{2\pi} (\theta_{i,c} - \theta_{j,c})\bigg)  - g \sum_{\langle i,j\rangle} \cos \bigg( \sqrt{2\pi} (\phi_{i,c} - \phi_{j,c}) \bigg), 
\label{quasi1d}
\end{align}
in which the sum $\langle i,j \rangle$ runs over the nearest neighboring ladders. This  Hamiltonian was investigated in [\onlinecite{jaefari-2010}], and has are three phases: SC, SC+ CDW, and CDW phases. 

In the SC phase, the PDW and $4e$ SC order parameters develop the two-dimensional order. On the boundary, there will be Majorana fermions zero modes  coming from the charge sectors (in analogy from the two PDW ladder case.) If there are the $N$ ladders, then there are $(N-1)$ Majorana zero modes at the boundary in this phase since there are $(N-1)$ cosine's in Eq.\eqref{quasi1d}. 

In the CDW phase, there is no SC order parameter with long-range order. Hence, there is no Majorana fermion from the charge sector in the CDW phase. In the SC+CDW phase, there may be Majorana fermions depending on the relative strength of the two order parameters. If the CDW order parameter is stronger than SC order parameter, then there is no Majorana fermion since each charge sector appearing in the sum of Eq.\eqref{quasi1d} is effectively in the CDW phase of Eq.\eqref{boskitaev}. If the SC order parameter is the strongest, then there will be Majorana fermions from the charge sectors. 

We now address the fate of the Majorana fermions from the spin sector. Because the charge sectors decouple from the spin sectors, the discussion here will apply to all the three phases of the charge sector. Though any inter-ladder spin-spin interaction is irrelevant in bulk, such term  make the boundary Majorana fermion $\gamma_{j, s}$ of the $j$-th ladder have an interaction with its nearest neighbors 
\begin{align}
\delta H' = - it \sum_{\langle i,j \rangle} \gamma_{i,s}\gamma_{j,s}.
\end{align}
The low-energy physics {\it at the boundary} is described by  
\begin{align}
\delta H' \to -iv_{F} (\gamma_{R,s}\partial_{x} \gamma_{R,s} - \gamma_{L,s}\partial_{x}\gamma_{L,s}), 
\end{align}
in which $(\gamma_{R}, \gamma_{L})$ are sitting at $k =0$ and $k= \pi$ in momentum space.\cite{cho-2012b} If  translation symmetry along $x$ is imposed, then the system has protected gapless Majorana fermions. If not, the Majorana fermions may dimerize and break the translational symmetry spontaneously.\cite{cho-2012b} We expect that the same analysis applies to the Majorana fermions from the charge sector. 

We now address if the topological defects of the various topological phases trap a Majorana fermion or not. Because the phases in hand are {\it weak} topological phases, the vortices do not carry any Majorana zero mode. Instead, a Majorana zero mode will be localized if there is a {\it lattice dislocation} with the Burgers vector {\em perpendicular} to the ladders.\cite{cho-2012b, ran-2010, teo-2013} As a byproduct, we notice that the quasi-1D SC state emerging from a Luther-Emery liquid\cite{kivelson-1998, emery-2000, jaefari-2010} (a spin-gapped Luttinger liquid), where the same form of the Hamiltonian was studied, also supports Majorana fermions on the boundary and at the lattice dislocations.